\documentclass[10pt,aps,prl,showpacs,twocolumn,longbibliography,groupedaddress]{revtex4-2}
\usepackage{graphicx}
\usepackage{latexsym}
\usepackage{amssymb}
\usepackage{bbold}
\usepackage{bm}
\usepackage{color}
\usepackage[pdftex,bookmarks,colorlinks,breaklinks]{hyperref} 
\hypersetup{linkcolor=red,citecolor=blue,filecolor=dullmagenta,urlcolor=blue} 
\hypersetup{pdftoolbar=true,pdfpagemode=UseNone,pdfstartview=FitH, colorlinks=true,extension=}

\newcommand{\onehalf}{$\frac{1}{2}$\,}
\newcommand{\unity}{\mathbb{1}}
\newcommand{\bi}[1]{\textbf{\textit{#1}}}
\newcommand{\bh}{\bi{h}}
\newcommand{\bG}{G}
\newcommand{\bS}{S}
\newcommand{\ba}{a}
\newcommand{\bb}{b}
\newcommand{\bphi}{\bm{\varphi}}

\begin{document}
\title{Spin resonance without a spin: A microwave analog}
\author{Tobias Hofmann}
\affiliation{Fachbereich Physik der Philipps-Universit\"{a}t Marburg, D-35032 Marburg, Germany, EU}
\author{Finn Schmidt}
\affiliation{Fachbereich Physik der Philipps-Universit\"{a}t Marburg, D-35032 Marburg, Germany, EU}
\author{Ulrich Kuhl}
\email{ulrich.kuhl@univ-cotedazur.fr}
\affiliation{Universit\'{e} C\^{o}te d'Azur, CNRS, Institut de Physique de Nice (INPHYNI), 06108 Nice, France, EU}
\affiliation{Fachbereich Physik der Philipps-Universit\"{a}t Marburg, D-35032 Marburg, Germany, EU}
\author{Hans-J\"urgen St\"{o}ckmann}
\email{stoeckmann@physik.uni-marburg.de}
\affiliation{Fachbereich Physik der Philipps-Universit\"{a}t Marburg, D-35032 Marburg, Germany, EU}

\date{\today}

\begin{abstract}
An analog of nuclear magnetic resonance is realized in a microwave network with symplectic symmetry.
The network consists of two identical subgraphs coupled by a pair of bonds with a length difference corresponding to a phase difference of $\pi$ for the waves traveling through the bonds.
As a consequence all eigenvalues appear as Kramers doublets.
Detuning the length difference from the $\pi$ condition Kramers degeneracy is lifted, which may be interpreted as a Zeeman splitting of a spin \onehalf in a magnetic field.
The lengths of another pair of bonds are modulated periodically with frequencies of some 10\,MHz by means of diodes, thus emulating a magnetic radiofrequency field.
Features well-known from NMR such as the transition from the laboratory to the rotating frame, and Lorentzian shaped resonance curves can thus be realized.
\end{abstract}

\maketitle

The statistical features of the spectrum of a quantum-mechanical system depend crucially on its properties with respect to time-reversal symmetry (TRS).
For systems with TRS, e.\,g., there is an antiunitary symmetry $T$ obeying $T^2=1$ if there is no spin \onehalf, and $T^2=-1$ in the presence of a spin \onehalf.
Joyner and coworkers \cite{joy14} noticed that a spin \onehalf is not really needed for the latter case, any system obeying a symmetry $T$ with $T^2=-1$ will do it as well.
They proposed a correspondingly designed graph, which was experimentally realized by us in a microwave network \cite{reh16,reh18}.
Here we take this analogy literally and ask the question:
If there is a spin analog in a network with an anti-unitary  symmetry obeying $T^2=-1$, is there, perhaps, also an analog of spin resonance?

We start with a recapitulation of the basic ideas of nuclear magnetic resonance \cite{sli80}:
A nuclear spin $\vec{I}$ is exposed to a static magnetic field $B_0$, by convention in $z$ direction, and to a radiofrequency field $B_1(t)=2B_1\cos(\omega_Rt)$ in $x$ direction, generated by a coil wrapped around the probe.
The system is described by the time-dependent Schr\"odinger equation
\begin{equation} \label{eq:schr1}
	\dot{\psi}=-\frac{i}{\hbar}H_\mathrm{NMR}\psi
\end{equation}
with the Hamiltonian
\begin{equation} \label{eq:HNMR}
	H_\mathrm{NMR}=-\hbar\left[\omega_0I_z+2\omega_1\cos(\omega_Rt) I_x\right]\,,
\end{equation}
where $\omega=\gamma B$, with the gyromagnetic ratio $\gamma$.
Assuming a spin $I$=\onehalf, the angular momentum operators can be expressed in terms of the Pauli matrices, $\vec{I}=\frac{1}{2}\vec{\sigma}$.

\begin{figure}
	\includegraphics[width=\linewidth]{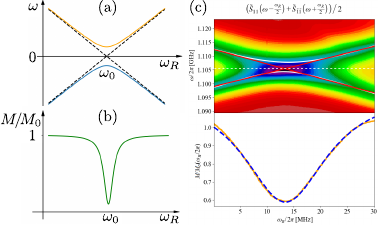}
	\caption{ \label{fig:NMR} 
		(a) Eigenvalues of a spin \onehalf system, exposed to a static field in $z$ and a radiofrequency field in $x$ direction, in the rotating frame in a classical NMR experiment, and 
		(b) magnetization in dependence of the Larmor angular frequency $\omega_R$ of the radiofrequency field.
		(c) ``NMR'' realization in a microwave network (see text for details).
	}
\end{figure}

The standard approach to solve the Sch\"odinger equation is a transformation into a rotating frame to remove the time-dependency.
With the rotated wave function,
\begin{equation} \label{eq:psiR}
	\psi_R=e^{-i\frac{\omega_R}{2}t\sigma_z}\psi
\end{equation}
the Schr\"odinger equation is transformed into
\begin{equation} \label{eq:schr2}
	\dot{\psi}_R=-\frac{i}{\hbar}H_\mathrm{NMR}^R\psi_R
\end{equation}
with
\begin{equation} \label{eq:HNMRR}
	H_\mathrm{NMR}^R= -\frac{1}{2}\left[(\omega_0-\omega_R)\sigma_z+\omega_1\sigma_x \right]\,,
\end{equation}
where terms rotating with $2\omega_R t$ have been discarded.
The eigenvalues of $H_\mathrm{NMR}^R$ are given by
\begin{equation} \label{eq:eigen}
	\omega_\pm=\pm\frac{1}{2}\sqrt{(\omega_0-\omega_R)^2+\omega_1^2}\,,
\end{equation}
see Fig.~\ref{fig:NMR}(a).
In a standard NMR experiment the magnetization $M$, proportional to the spin polarization, is studied as a function of $\omega_0$ or $\omega_R$.
The avoided crossing exhibited by the eigenvalues in the rotating frame at $\omega_0=\omega_R$ implies a Lorentzian resonance curve,
\begin{equation} \label{eq:res}
	M/M_0=1-\frac{\omega_1^2} {(\omega_0-\omega_R)^2 +\omega_1^2}\,,
\end{equation}
see Fig.~\ref{fig:NMR}(b), where $M_0$ is the equilibrium polarization, usually resulting from a Boltzmann polarization.
We shall follow exactly the same strategy in our approach to realize an NMR analogue in a microwave network.

\begin{figure}
	\includegraphics[width=0.9\linewidth]{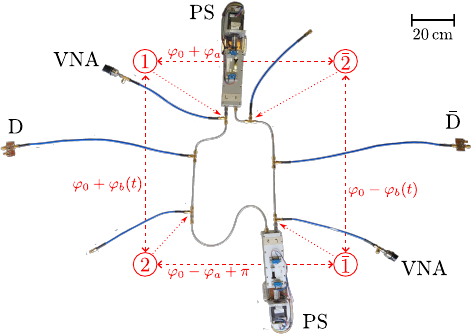}\\
	\caption{ \label{fig:photo_setup}
		Photo of the experimental set-up.
		The graph is connected to a vector network analyzer (VNA) via two cables attached at nodes $1$ and $\bar{1}$.
		Two phase shifters (PS) allow for a change of the lengths of bonds $1\bar{2}$ and $\bar{1}2$.
		The $\pi$ jump is realized by an appropriate additional length of bond $\bar{1}2$.
        By means of diodes (D) and ($\bar{\mathrm{D}}$), attached at the ends of two dangling bonds, the effective lengths of bonds $12$ and $\bar{1}\bar{2}$ can be switched between two states with frequencies up to 125\,MHz.
		In addition the phase shifts are given which waves acquire when traveling through the respective bonds.
    }
\end{figure}

Figure~\ref{fig:photo_setup}(a) shows a photo of the studied graph.
The unperturbed graph consists of four nodes $1,2,\bar{1}, \bar{2}$, coupled by four bonds of equal length $l$.
At each of the nodes a dangling bond is attached, again of length $l$, and terminated by short ends denoted by $1_D$, $\bar{1}_D$, $2_D$, $\bar{2}_D$.
At $1_D$ and $\bar{1}_D$ the graph is weakly coupled to a vector network analyzer, achieved by T junctions at the coupling points with short terminated side ports.
A phase shift of an odd integer multiple of $\pi$, needed for the symplectic symmetry, is produced by an additional length of bond $\bar{1}2$ \cite{reh16}.
The static length shifts are performed by phase shifters, microwave standard devices varying the phase by changing the length.
By means of diodes, attached at the end of dangling bonds, the effective lengths of bonds $12$ and $\bar{1}\bar{2}$ can be switched between two states with frequencies up to 125\,MHz.
The lengths of the bonds thereby are not varied explicitly, but the phase shift a wave experiences when traveling through the bonds.

The bonds are dielectric circular waveguides with a teflon dielectric ($n= 1.44$) separating the central core from the outer cylinder.
For frequencies of some GHz as applied in the present work the wave guides support one traveling mode only \cite{hul04}.
The waves $\psi_{nm}(x,t)$ in the bonds $nm$ obey the time-dependent wave equation.
In the calculations variations of the length $l$ would be inconvenient.
But since the $\psi_{nm}(x,t)$ depend on the {\em optical} length $l=l_\mathrm{opt}=nl_\mathrm{geo}$ only, a change of the {\em geometrical} length $l_\mathrm{geo}$ may be substituted by a corresponding change of $n$.
The $\psi_{nm}(x,t)$ then are solutions of a correspondingly rescaled wave equation,
\begin{equation} \label{eq:wave}
	\left[-\frac{n^2}{c^2}\frac{\partial^2}{\partial t^2}+\frac{\partial^2}{\partial x^2}+2v_{nm}\frac{\partial^2}{\partial x^2}\right]\psi_{nm}(x,t)=0\,,
\end{equation}
where $v_{nm}=0$ for the dangling bonds, $v_{1\bar{2}/\bar{1}2}= \pm a$, and $v_{12/\bar{1}\bar{2}}= \pm b(t)$.
$a$ and $b(t)=2b\cos(\omega_Rt)$ are the relative length changes due to phaseshifter and diode, respectively.
It has been assumed that these changes are small and can be treated in first order.

Continuity of the wave functions at the vertices and current conservation result in a secular equation \cite{kot99a},
\begin{equation} \label{eq:h}
	\bh\bm{\varphi}=0\,,
\end{equation}
for $\bm{\varphi}$, the vector of the voltages $\varphi_n$ at the vertices.
$\bh$ is the secular matrix \cite{reh16,reh18}, in the present case given by
\begin{equation} \label{eq:sec2}
	\bh=\bh_0+\bh_a+\bh_b\,,
\end{equation}
where, with the sequence $1,\bar{1},2,\bar{2}$ of rows and columns,
\begin{equation} \label{eq:H0}
	\bh_0 = \left(
		\begin{array}{cc}
			-3f\, \unity & g(\unity+i\sigma_y) \\
			g(\unity-i\sigma_y) & -3f\, \unity \\
		\end{array}
	\right)\,,
\end{equation}
is the secular matrix for the unperturbed graph, with $f=k\cot(kl)$, $g=k/\sin(kl)$, and
\begin{equation}
	\bh_a = ak\left(
		\begin{array}{cc}
			f'\sigma_z & -g'\sigma_x \\
			-g'\sigma_x & -f'\sigma_z \\
		\end{array}
	\right)\,, 
	\bh_b=bk\left(
		\begin{array}{cc}
			f'\sigma_z &-g'\sigma_z \\
			-g'\sigma_z & f'\sigma_z \\
		\end{array}
	\right)
\end{equation}
are the contributions due to the perturbation, where $f'$, $g'$ denote the derivatives of $f$ and $g$ with respect to $k$.

Now we turn to the basis where $\bh_0$ is diagonal.
This is achieved by means of the transformation
\begin{equation} \label{eq:trans}
	\tilde{\bphi} =
		\frac{1}{\sqrt{2}}\left(
			\begin{array}{cc}
				\varepsilon_y & i\sigma_y \varepsilon_y^*\\
				i\sigma_y\varepsilon_y & \varepsilon_y^* \\
			\end{array}
		\right) \bphi
	\,,\quad
	\varepsilon_y=e^{i\frac{\pi}{8}\sigma_y}\,.
\end{equation}
The transformed $\tilde{\bh}_0$ is diagonal with an upper block $\tilde{\bh}_{0U}=(-3f+g\sqrt{2})\cdot \unity$, and a lower one $\tilde{\bh}_{0D}=(-3f-g\sqrt{2})\cdot \unity$, illustrating symplectic symmetry and Kramers degeneracy.

For the measurement the graph is connected via vertices $1$ and $\bar{1}$ to a vector network analyzer measuring reflection amplitudes $S_{11}$, $S_{\bar{1}\bar{1}}$ and transmission amplitudes $S_{1\bar{1}}$, $S_{\bar{1}1}$ between the ports.
The scattering matrix
\begin{equation} \label{eq:S}
	\bS={\left(
		\begin{array}{cc}
			S_{11} & S_{1\bar{1}} \\
			S_{\bar{1}1} &S_{\bar{1}\bar{1}}
		\end{array}
	\right)}
\end{equation}
relates vectors $\ba=(a_1,a_{\bar{1}})^T$, $\bb=(b_1,b_{\bar{1}})^T$ of incoming and outgoing amplitudes, respectively, via
\begin{equation} \label{eq:scatt}
	\bb=\bS\ba\,.
\end{equation}
In the eigenbasis of $\bh_0$ $\ba$ and $\bb$ are transformed for the upper block into $\tilde{\ba}= \frac{1}{\sqrt{2}}\varepsilon_y \ba$ and $\tilde{\bb}= \frac{1}{\sqrt{2}}\varepsilon_y \bb$.

Let us first discuss the static case, $b(t)=b$.
Here we can apply time-independent scattering theory \cite{guh98} establishing a relation between the scattering matrix $\bS$ and the Green function $\bG$
\begin{equation} \label{eq:S0}
	\bS=\unity-\frac{2i\gamma\bG}{\unity+i\gamma\bG}\,,
\end{equation}
where $\gamma$ contains the information on the antenna coupling.
In the present case with only one pair of antennas at symmetry equivalent points $\bS$ and $\bG$ are $2\times 2$ matrices.
$\bG$ is related to the secular matrix via \cite{kot99a}
\begin{equation} \label{eq:Gh}
	\bG=\left(
		\begin{array}{cc}
			(\bh^{-1})_{11} & (\bh^{-1})_{1\bar{1}} \\
			(\bh^{-1})_{\bar{1}1} &(\bh^{-1})_{\bar{1}\bar{1}}
		\end{array}
	\right)\,.
\end{equation}
In the case of symplectic symmetry $\bG$ exhibits poles at the Kramers doublets.
Expressing $\bG$ as a sum over the poles, and restricting the discussion to the neighborhood of just one Kramers doublet, Eq.~(\ref{eq:S0}) for symplectic symmetry simplifies to
\begin{equation} \label{eq:S1}
	\bS=\unity-\frac{2i\gamma'}{(\omega+i\gamma')\unity -H}\,,\quad H=\omega_n\unity\,,
\end{equation}
with $\gamma'=\gamma g_n$, where $\omega_n$ and $g_n$ are position and residuum of the selected Kramers doublet, respectively.

\begin{figure}
	\includegraphics[width=.99\linewidth, height=9cm]{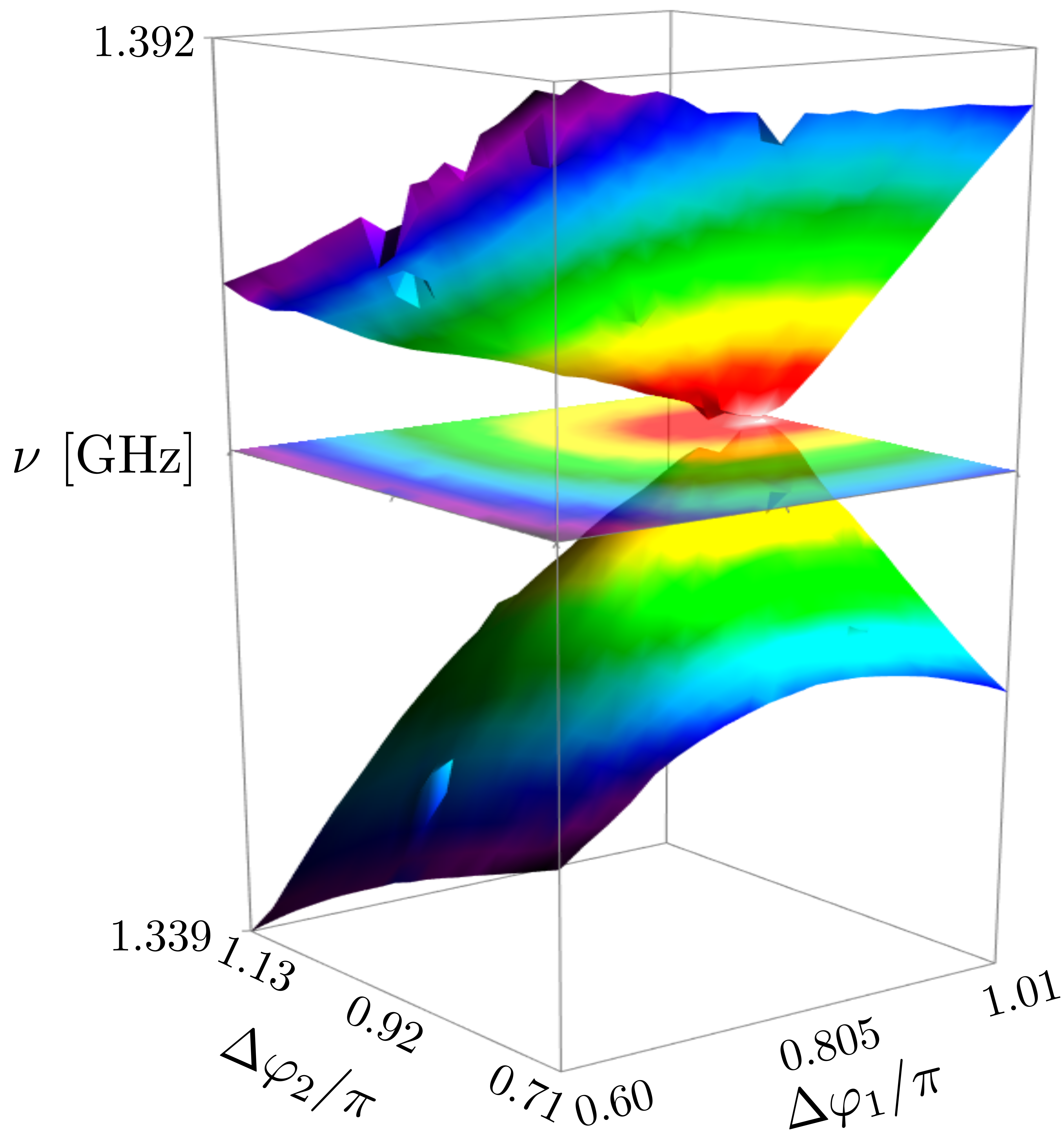}\\
	\caption{ \label{fig:diabolo}
		Eigenfrequencies of the split Kramers doublet at $\omega/2\pi = \nu=1.391$\,GHz in a graph, composed two subgraphs, complex-conjugates of each other, with two pairs of coupling bonds to simulate the magnetic fields, in dependence of the phase differences $\Delta\varphi_1=k\Delta l_1$ and $\Delta\varphi_2=k\Delta l_2$.
        In addition the projection of the diabolo onto the $\Delta\varphi_1$-$\Delta\varphi_2$ plane is shown.
	}
\end{figure}

\begin{figure*}
	\includegraphics[width=0.9\linewidth]{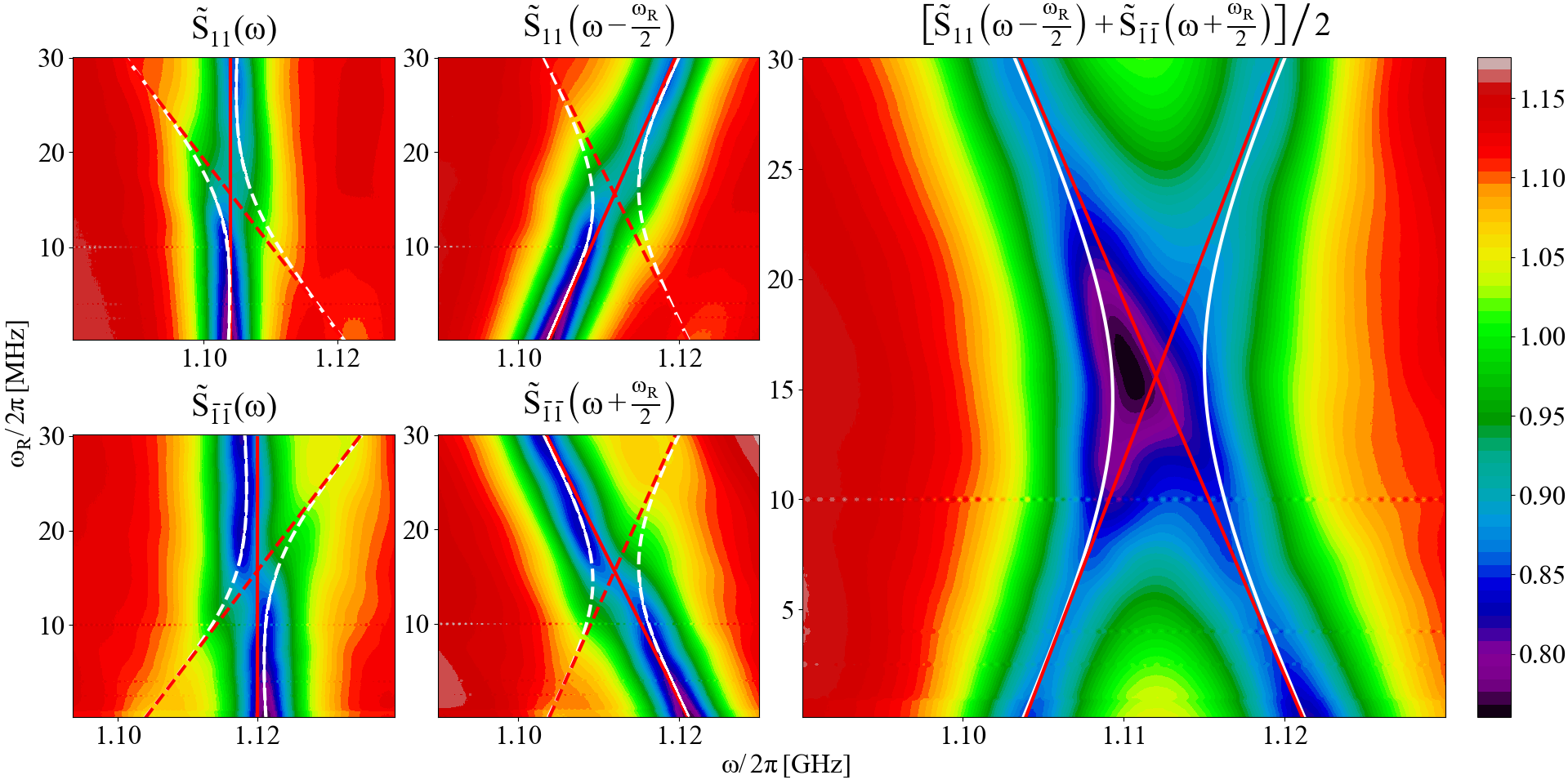}\\
	\caption{ \label{fig:Srot}
		From the laboratory to the rotating frame (see text for details).
		}
\end{figure*}

As soon as the symplectic symmetry is broken, the contributions to $h$ from the perturbation generate two extra terms to $H$,
\begin{equation} \label{eq:HNMR0}
	H = \omega_n\unity-\frac{\omega_1}{2}\sigma_z - \frac{\omega_2}{2}\sigma_x\,,
\end{equation}
where $\omega_1\sim a$, and $\omega_2\sim b$.
Details of the calculation will be presented elsewhere \cite{sch24}.
The notation has been chosen to be in accordance with NMR practice, see Eq.~(\ref{eq:HNMR}).
We have thus established the equivalence of the splitting of a Kramers doublet by the perturbations with the Zeeman splitting of a spin \onehalf by two magnetic fields in $z$ and $x$ directions.

The eigenvalues of the Hamiltonian~(\ref{eq:HNMR0}) are given by
\begin{equation} \label{eq:zeeman}
	\omega_\pm=\omega_n\pm\frac{1}{2}\sqrt{\omega_1^2+\omega_2^2}\,.
\end{equation}
Figure~(\ref{fig:diabolo}) shows the results from an earlier study, where the two fields had been realized by two pairs of coupling bonds \cite{reh18}.
The eigenvalues are plotted in dependence of $\Delta\varphi_1=k\Delta l_1$ and $\Delta\varphi_2=k\Delta l_2$, the phase differences the two waves acquire when traveling through the pairs of bonds.
In the three-dimensional space with the $\omega$ axis perpendicular to the $\Delta\varphi_1$-$\Delta\varphi_2$ plane the eigenvalues~(\ref{eq:zeeman}) form a double cone, a {\em diabolo}, see Fig.~\ref{fig:diabolo}.
At the diabolic point $\Delta\varphi_1=\Delta\varphi_2=\pi$, corresponding to $\omega_1=\omega_2=0$ the graph is symplectically symmetric.

Let us now move to the discussion of the time-dependent case.
Guided by the procedure applied in classical NMR we look for a transformation $\varphi_R=R(t)\varphi$ into a rotating frame which removes the time dependence in the wave equation~(\ref{eq:wave}).
The details are presented in \cite{sch24}.
Here it is sufficient to know that the amplitudes in the rotating frame
$\ba_R(\omega,t)=e^{-i\omega t}\ba_R(\omega)$ and $\bb_R(\omega,t)=e^{-i\omega t}\bb_R(\omega)$ are related to the corresponding quantities in the laboratory frame, $\tilde{\ba}(\omega,t)$ and $\tilde{\bb}(\omega,t)$ via spinor rotations,
\begin{equation} \label{eq:aR}
  \tilde{\ba}(\omega,t) 
    = e^{i\frac{\omega_Rt}{2}\sigma_z} \ba_R(\omega,t)
    = \left(\begin{array}{c}
      e^{-i\,\omega_-t}\,(\ba_R)_1(\omega) \\
      e^{-i\,\omega_+t}\,(\ba_R)_{\bar{1}}(\omega)\\
      \end{array}\right)\,,
\end{equation}
with $\omega_{\pm}=\omega\pm\frac{\omega_R}{2}$.
A corresponding formula holds for $\tilde{\bb}(\omega,t)$.
By this transformation the time dependence in the wave equation has disappeared, apart from a term rotating with $2\omega_Rt$, which is neglected.
An additional term results from the transformation of the time derivative in the wave equation.
Note that a completely analogous spinor rotation had been applied to remove the time dependence of the NMR Hamiltonian~(\ref{eq:HNMR}), see Eq.~(\ref{eq:psiR}).

Equation~(\ref{eq:scatt}) is transformed by the rotation into
\begin{equation} \label{eq:scattR}
	\bb_R=\bS_R\ba_R\,.
\end{equation}
Equation~(\ref{eq:S1}) also holds for $\bS_R$, but with $H$ replaced by
\begin{equation} \label{eq:Gq}
	H_R=\omega_n\unity+ H^R_\mathrm{NMR}\,,
\end{equation}
where $H^R_\mathrm{NMR}$ is exactly the NMR Hamiltonian~(\ref{eq:HNMRR}) in the rotating frame.
Equations~(\ref{eq:aR}) and (\ref{eq:scattR}) show that the upper left and lower right matrix elements of $\bS_R(\omega)$ may be interpreted as
\begin{equation} \label{eq:SR}
	(\bS_R)_{11}(\omega)=\tilde{S}_{11}(\omega_-)\,,\quad
	(\bS_R)_{\bar{1}\bar{1}}(\omega)=\tilde{S}_{\bar{1}\bar{1}}(\omega_+)\,.
\end{equation}
The off-diagonal elements of $\bS_R$ cannot such be determined,
since this would mean an excitation of the graph at the frequency $\omega_+$, and a detection at $\omega_-$, and vice versa (possible in principle, but not with the available equipment).
But the trace of $\bS_R(\omega)$,
\begin{equation} \label{eq:TrSR}
	\mathrm{Tr}\bS_R(\omega)=\tilde{S}_{11}(\omega_-)+\tilde{S}_{\bar{1}\bar{1}}(\omega_+)\,,
\end{equation}
is accessible.
The dependency of $\bS_R(\omega)$ on $\omega_R$ has not been noted explicitly to simplify the notation.
Equation~(\ref{eq:TrSR}) shows that a {\em twisted} excitation is needed to get the scattering matrix in the rotating frame:
The spin-up component has to be excited at $\omega_-$, the spin-down component at $\omega_+$, a really strange situation.

This is illustrated for a Zeeman split Kramers doublet with components at frequencies $\nu_1=\omega_1/2\pi= 1.105$\,GHz and $\nu_2=\omega_2/2\pi= 1.120$\,GHz , see Fig.~\ref{fig:Srot}.
The left column shows $\tilde{S}_{11}(\omega)$ (top) and $\tilde{S}_{\bar{1}\bar{1}}(\omega)$ (bottom) as a function of frequency $\nu=\omega/2\pi$ and rotation frequency $\nu_R=\omega_R/2\pi$ in a color plot.
$\tilde{S}_{11}(\omega)$ sees only the lower frequency component of the Kramers doublet, and is blind for the other one.
For $\tilde{S}_{\bar{1}\bar{1}}(\omega)$ it is vice versa.
Both $\tilde{S}_{11}(\omega)$ and $\tilde{S}_{\bar{1}\bar{1}}(\omega)$ are nearly independent of $\omega_R$.
Only at $\nu_R=\omega_R/2\pi= 15$\, MHz, corresponding to the splitting $\Delta\omega$ of the two Kramers doublets, there is an indication that something is happening.
The central column of $\nu_R=\omega_R/2\pi$ shows the same data, but with $\tilde{S}_{11}$ and $\tilde{S}_{\bar{1}\bar{1}}$ plotted in dependence of the twisted frequencies $\omega_-$ and $\omega_+$, respectively.
The figure on the right finally shows the sum of the two latter quantities corresponding to $\mathrm{Tr}\bS_R(\omega)$, see Eq.~(\ref{eq:TrSR}).
Now two hyperbolic branches become clearly visible, corresponding to the eigenvalues in the rotating frame, and exhibiting an anti-crossing at the resonance position $\omega_R=\Delta\omega$.

By means of this somewhat tricky operation we have been able to convert the measured $S$ matrix components $\tilde{S}_{11}(\omega)$ and $\tilde{S}_{\bar{1}\bar{1}}(\omega)$ in the laboratory frame into the trace of the $S$ matrix $\bS_R(\omega)$ in the rotating frame.
As a result we have obtained the spectrum of the NMR Hamiltonian in the rotating frame, see Eq.~(\ref{eq:eigen}), a quantity not even available in a standard NMR experiment.

Of particular interest is the behavior of $\mathrm{Tr}\bS_R(\omega)$ along the vertical symmetry line defined by $\omega=0$,
\begin{equation} \label{eq:res1}
	M = \frac{1}{2} \mathrm{Tr}\bS_R(0)=1-\frac{4\gamma'^2}{(\omega_0-\omega_R)^2 +\omega_1^2+(2\gamma')^2}\,,
\end{equation}
see Eq.~(\ref{eq:S1}), with $\bS$ and $H$ replaced by $\bS_R$ and $H_R$, respectively.
This is exactly the expression for a typical Lorentzian shaped magnetic resonance curve, see Eq.~(\ref{eq:res}), with the only difference that now there is an additional contribution to the resonance width from the coupling.
The right column of Fig.~\ref{fig:NMR} illustrates this for the same Kramers doublet discussed above.
The upper figure shows again the right part of Fig.~\ref{fig:Srot}, but with interchanged axes.
The lower figure shows the corresponding Lorentzian ``resonance curve''.
Comparison with the left column of Fig.~\ref{fig:NMR} shows the complete correspondence with the behavior found in classical NMR.

Is this strange realization of a spin resonance of any use, or is it just a crazy idea? One perspective crossing immediately one's mind are spin relaxation studies.
In standard NMR relaxation measurements are performed to get information on the origin of the fluctuating interactions.
Here it is vice versa:
It is easy to generate well-controlled fluctuating interactions, again by means of diodes, and to study their implications for the resonance line shape.
A prerequisite for this achievement was the development of a technique allowing for rapid changes of the transmission properties of networks with frequencies up to 125\,MHz.
Thereby a completely new class of systems, time-dependent graphs, in particular Floquet systems, has become accessible to the experiment.

One of the authors (H.-J.~St.) thanks the department of physics of the university of Marburg for providing him with every support needed to continue with his research over many years after his official retirement.

\end{document}